\documentclass[
reprint,
superscriptaddress,
amsmath,
amssymb,
prl,
floatfix,
longbibliography
]{revtex4-1}

\usepackage{graphicx}
\usepackage{dcolumn}
\usepackage{bm}
\usepackage[colorlinks=true]{hyperref}
\usepackage{url}

\usepackage{color}
\usepackage{braket}
\usepackage{bm}
\usepackage{textcomp}

\newcommand{\gatename}[1]{\ensuremath{\mathsf{#1}}}

\let\transpose\intercal

\begin{document}

\title{Quantum Kitchen Sinks: An algorithm for machine learning on near-term quantum computers}

\author{C.\ M.\ Wilson}
\affiliation{Rigetti Computing, 2919 Seventh Street, Berkeley, CA, 94710-2704 USA}
\affiliation{Institute for Quantum Computing, University of Waterloo, Waterloo, N2L 3G1, Canada}
\affiliation{Department of Electrical and Computer Engineering, University of Waterloo, Waterloo, N2L 3G1, Canada}
\author{J.\ S.\ Otterbach}
\altaffiliation[Current address: ]{OpenAI}
\affiliation{Rigetti Computing, 2919 Seventh Street, Berkeley, CA, 94710-2704 USA}
\author{N.\ Tezak}
\altaffiliation[Current address: ]{OpenAI}
\affiliation{Rigetti Computing, 2919 Seventh Street, Berkeley, CA, 94710-2704 USA}
\author{R.\ S.\ Smith}
\affiliation{Rigetti Computing, 2919 Seventh Street, Berkeley, CA, 94710-2704 USA}
\author{A.\ M.\ Polloreno}
\altaffiliation[Current address: ]{University of Colorado, Boulder, CO}
\affiliation{Rigetti Computing, 2919 Seventh Street, Berkeley, CA, 94710-2704 USA}
\author{Peter\ J.\ Karalekas}
\affiliation{Rigetti Computing, 2919 Seventh Street, Berkeley, CA, 94710-2704 USA}
\author{S.\ Heidel}
\altaffiliation[Current address: ]{Airbnb}
\affiliation{Rigetti Computing, 2919 Seventh Street, Berkeley, CA, 94710-2704 USA}
\author{M.\ Sohaib\ Alam}
\affiliation{Rigetti Computing, 2919 Seventh Street, Berkeley, CA, 94710-2704 USA}
\author{G.\ E.\ Crooks}
\altaffiliation[Current address: ]{Google X}
\affiliation{Rigetti Computing, 2919 Seventh Street, Berkeley, CA, 94710-2704 USA}
\author{M.\ P.\ da Silva}
\altaffiliation[Current address: ]{Quantum Systems Group, Microsoft, One Microsoft Way, Redmond, WA 98052}
\email{marcus.silva@microsoft.com}
\affiliation{Rigetti Computing, 2919 Seventh Street, Berkeley, CA, 94710-2704 USA}
\date{\today}

\begin{abstract}
  Noisy intermediate-scale quantum computing devices are an exciting platform
  for the exploration of the power of near-term quantum applications.
  Performing nontrivial tasks in such devices requires a
  fundamentally different approach than what would be used on an error-corrected 
  quantum computer. One such approach is to use {\em hybrid
    algorithms}, where problems are reduced to a parameterized quantum
  circuit that is often optimized in a classical feedback loop. Here we
  describe one such hybrid algorithm for machine learning tasks by
  building upon the classical algorithm known as {\em random kitchen sinks}. Our
  technique, called {\em quantum kitchen sinks}, uses quantum circuits
  to nonlinearly transform classical inputs into features that
  can then be used in a number of machine learning algorithms.
  We demonstrate the power and flexibility of this proposal by using it to solve
  binary classification problems for synthetic datasets as well as handwritten
  digits from the MNIST database. Using the Rigetti quantum virtual machine, we show that small quantum circuits provide significant performance lift over standard
  linear classical algorithms, reducing classification error rates from
  50\% to $<0.1\%$, and from $4.1\%$ to $1.4\%$ in these two examples, respectively. Further, we are able to run the MNIST classification problem, using full-sized MNIST images, on a Rigetti quantum processing unit, finding a modest performance lift over the linear baseline.
\end{abstract}

\maketitle

\paragraph{\label{intro}Introduction---} Interest in
adapting or developing machine learning algorithms for near-term
quantum computers has grown rapidly.  While quantum machine learning
(QML) algorithms offering exponential speed-ups on universal quantum
computers have been known for some
time~\cite{HHL2009,Reben2014,LMR2014,CSS2015,KP2016,BS2017,BKL+2017},
recent interest has increasingly focused on algorithms for noisy,
intermediate-scale quantum (NISQ) computers~\cite{Pre2018,
  Schuld2018_2, Grant2018, Havlicek2018, Huggins2018}. These
algorithms aim to minimize the complexity of the required quantum
circuit so that they may be executed by NISQ devices while still
yielding meaningful results.  This is in contrast to approaches that
allow for arbitrarily large circuits of width and depth that grow
polynomially in the input size. These approaches can only yield meaningful
answers if errors are suppressed to rates that are inversely
proportional to the circuit size, something that is not possible with
NISQ devices and requires fault tolerance~\cite{KLZ1998,AGP2006,ABO2008}.

Many of the proposed approaches use a so-called hybrid model for NISQ
computing, where the quantum processor is considered an expensive
resource and is extensively supported by classical
computing resources.  In particular, many of these proposals use a
variational approach, where parameters of a small quantum circuit are
optimized using classical optimization algorithms which use
measurement outcomes to compute a cost
function~\cite{PMS+2014,FGG2014,KMT+2017,FGGN2017,YRS+2017,FN2018,Havlicek2018}.
While these closed-loop hybrid approaches move the computational cost
of the optimization algorithm off of the quantum hardware, the
iterative nature of the optimization process still requires a large
number of calls to the ``expensive'' quantum resource.

In this paper, we propose a QML algorithm that eliminates the need
for costly parameter optimization of quantum circuits.  This novel
open-loop hybrid algorithm, which we call {\em quantum kitchen sinks} (QKS), is
inspired by a technique known as {\em random kitchen sinks} whereby
random nonlinear transformations can greatly simplify the optimization
of machine-learning (ML) tasks~\cite{RR2008,RR2008b,RR2009}. The general
idea of QKS is to randomly sample from a family of quantum circuits
and use each circuit to realize a nonlinear transformation of the input data
to a measured bitstring. Subsequently, the concatenated results are processed
with a classical machine learning (ML) algorithm. This
approach is simple, flexible, and allows us to demonstrate that even
small quantum circuits, deep in the NISQ regime, can provide significant
``lift'' for complex ML tasks such as the classification of
hand-written digits. We further relate our circuits to common tools in ML
known as kernels.

\begin{figure}[t]
\includegraphics[width=0.6\columnwidth]{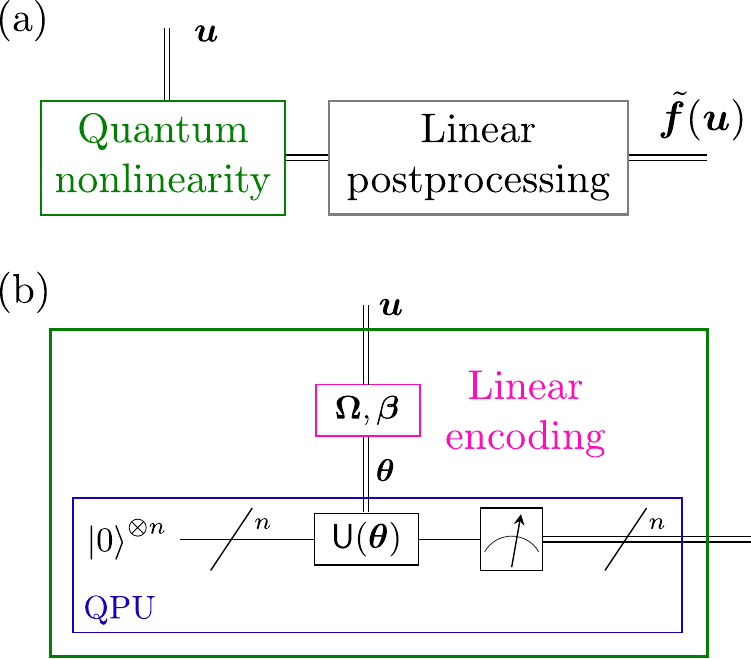}
\caption{
  (a)~Quantum kitchen sinks approximate a function $f$ applied to
  classical data $\mathbf{u}$ by using quantum circuits to apply a
  nonlinear transformation to $\mathbf{u}$ before additional
  classical (linear) postprocessing. (b)~The classical data are transformed
  by first encoding them into control parameters of a quantum circuit, and
  then measuring the quantum states. The results of measuring many
  different circuits parameterized by the same classical data are then
  collected into a single, large feature vector. \label{fig:data_flow}
}
\end{figure}

\paragraph{Random Kitchen Sinks---} The objective in supervised ML is
to approximate some \textit{a priori} unknown function $f(\bm u)$. For example,
this function may be a map from images, represented by the variable
$\bm u$, to labels, such as ``cat'' and ``dog''.  This is often done by
optimizing a parameterized function $g(\bm u; \bm \theta)$ to maximize
performance on a {\em training set} consisting of $M$ examples
$\{y_{i,\text{train}},\mathbf{u}_{i,\text{train}}\}$ such that
$g(\mathbf{u}_{i,\text{train}}; \bm \theta)\approx y_{i,\text{train}}$ for as many
examples in the training set as possible. The quality of the
approximation is often further quantified by how well $g$ performs on
some {\em test set} that is different from the training
set~\footnote{The use of a test set helps to avoid so-called
  ``overfitting'' of $g$ to the training set, as one would like an
  approximation to $f$ that generalizes well to previously unseen
  data, not simply an approximation that works only on the training
  set.}. A choice for $g$ that performs particularly well is a deep
neural network, which is a parametrized composition of many simple 
nonlinear functions, such as sigmoid functions or rectified linear units~\cite{Goodfellow2016}. 
Finding the parameters of $g$
that optimize performance (a process that for deep neural networks
is known as {\em deep learning}) can be resource intensive, requiring
large training sets and computational power~\cite{Goodfellow2016}.

Rahimi and Recht~\cite{RR2008,RR2008b,RR2009} observed that the costly
optimization of the training process could be replaced by
randomization.  In an approach dubbed {\em random kitchen sinks}
(RKS)~\cite{RR2008,RR2008b,RR2009}, they showed it was possible to
represent $g$ as a weighted, linear sum of simple nonlinear functions
that each have {\em random} parameters. Each term in this sum is called a
``kitchen sink''.  The weights of the sum still need to be optimized,
but this is a linear problem and, therefore, easy to solve. It has
been shown that, for example, the cosine, sign (i.e.,
$d\vert x \vert/dx$), and indicator functions can be used to obtain
good function approximations~\cite{RR2008}. The RKS idea originated from an
attempt to approximate the ``kernel trick''~\cite{SS2002,HSS2008}, by
randomly sampling eigenfunctions of an integration kernel. Since then,
this technique has been shown to apply beyond the sampling of a kernel, and to deliver performance that is
comparable to deep learning, while relying on much simpler numerical
techniques~\cite{MBL+2017,RR2017}.

The performance of the algorithm is dependent on the number of kitchen
sinks $D$, the choice of nonlinear function, and the number of
training examples $M$. Rahimi and Recht showed the approximation
error of $g$ in RKS scales as
$O(\frac{1}{\sqrt{D}}+\frac{1}{\sqrt{M}})$~\cite{RR2009} such that it may be
necessary to have large training sets and to generate many RKS in
order to achieve the same error rate as standard kernel
methods~\footnote{This means, in particular, that if the kernel
  corresponding to a circuit Ansatz is known, it is possible to estimate 
  the RKS error rate for that Ansatz by using a
  classical kernel machine, although performance as a function of $D$ is 
  often better than what would be expected from these bounds~\cite{RR2017}.}.

\paragraph{Classical vs.~quantum power---} Before discussing how to
generalize RKS to a quantum setting, we would like to make an
important observation.  In proposing an ML algorithm for quantum
computers, there is a danger that the quantum processor will not
contribute in a meaningful way to the power of the technique. If the
external classical part is powerful enough, the algorithm may work
\textit{in spite of} the transformation made by the quantum processor.
This can be seen as the flip-side of the RKS result we adapt: generic
nonlinearities in the {\em classical} processing can add power to the
ML algorithm, even if the quantum processing does not. For this
reason, it is important in a research context that the classical
portion of the algorithm be as simple and linear as possible. For this
reason, we will require all classical pre- and postprocessing to be
strictly linear, and consider only the added power of a nonlinear
transformation enabled by the quantum processor. We will refer to this
as the {\em Linear Baseline (LB) Rule}.

Applying the LB Rule to our strategy for testing and validation, 
we design an algorithm such that the quantum processor can
be removed and the input data can be passed directly through the
remaining (linear) classical part of the algorithm. We can then benchmark the
performance lift provided by the quantum processor against the
performance of the classical algorithm on its own. Note that a lift
provided by the quantum processor in this context does not imply an
absolute quantum advantage, but it does gives us a simple, operational
method to identify the power added by the quantum circuit.

\paragraph{Quantum Kitchen Sinks---}
We now describe our approach to translate the RKS framework
into something that may be computed by a quantum computer---what
we call QKS (see Fig.~\ref{fig:data_flow}).

\begin{figure}
  \includegraphics[width=0.45\textwidth]{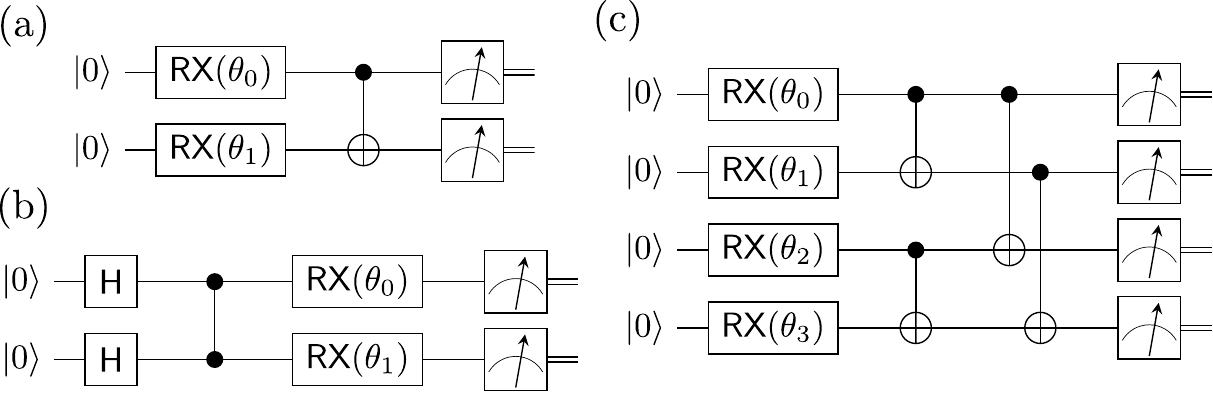}
  \caption{QKS Ans\"atze for (a)~two qubits using a \gatename{CNOT}, (b)~two
    qubits using a \gatename{CZ}, and (c) four qubits. Circuit (a) and
    (b) are interesting to contrast because (a) leads to high
    performance classification in multiple datasets, while (b) leads
    to classifiers that are no better than random (see the text for an
    explanation). Larger circuits are described in the
    appendix. \label{fig:2q_cnot}}
\end{figure}

As noted above, one of the nonlinear functions used to build RKSs is a
cosine. We can easily generate cosine transformations in a quantum setting 
by applying a Rabi rotation to a single qubit with a rate and phase that is 
chosen at random, but a time
duration that is a function of the input data. While this quantum
construction of RKSs works well, it can easily be simulated on a
classical computer.

In order to generalize this to circuits that are harder to simulate, our
first step is to specify the input data encoding in more detail. We
choose to encode the data into angles of rotations in the quantum
circuit, while keeping the state preparation and measurement fixed (a
similar approach was taken in~\cite{Havlicek2018} for a different QML
technique). This naturally leads the measurement statistics to depend
nonlinearly on the classical data.

Under the LB rule, we require that the mapping from data to angles be linear.
To define a linear encoding, let
$\mathbf{u}_i \in \mathbb{R}^p$ for $i = 1, \ldots, M$ be a $p$-dimensional
input vector from a data set containing $M$ examples. We can encode
this input vector into $q$ gate parameters using a $(q\times p)$-dimensional
matrix $\mathbf{\Omega}_e$ of the form
$\mathbf{\Omega}_e = (\mathbf{\omega}_1, \ldots, \mathbf{\omega}_q)^{\transpose}$
where $\mathbf{\omega}_k$ is a $p$-dimensional vector with a number
$r \le p$ elements being random values and the other elements being
exactly zero.  We can also specify a random $q$-dimensional bias
vector $\pmb\beta_e$. We then get our set of random parameters
$\bm\theta_{i,e}$ from the linear transformation
$\bm\theta_{i,e} = \mathbf{\Omega}_e \mathbf{u}_i + \pmb\beta_e$.
Notice the additional index $e$ which denotes the $e$\textsuperscript{th} episode,
i.e., the $e$\textsuperscript{th} repetition of the circuit parameterized through the
encoding $\mathbf{\Omega}_e, \pmb\beta_e$ (see below for a discussion
about episodes).

By specifying different elements of $\mathbf{\omega}_k$ to be nonzero,
we can specify different encodings. For instance, we can encode a
$p$-dimensional input vector into a single-qubit circuit by choosing
$q=1$ and $r=p$. In this single-qubit encoding, all dimensions of
$\mathbf{u}_i$ are combined into a single control parameter.
Conversely, we could use a split encoding with $q=p$ and $r=1$, where
each dimension of $\mathbf{u}_i$ is fed into a distinct control
parameter.  We discuss other possibilities below. Note that the set of
encoding parameters $\{\mathbf{\Omega}_e, \pmb\beta_e\}_{e=1}^{E}$ is
only drawn once and becomes a static part of the machine, which is
used for both training and testing. For the results presented in this
paper, the nonzero elements of $\mathbf{\Omega}_e$ are drawn from a
zero-mean normal distribution with variance $\sigma^2$, i.e.,
$\mathcal{N}(0,\sigma^2)$ and the elements of $\pmb\beta_e$ are drawn
from a uniform distribution $\mathcal{U}(0, 2\pi)$~\footnote{Note that
  the hyperparameter $\sigma$ must be optimized during training.}. However, other
distributions may also be considered. These choices only partially
determine the encoding. The exact structure of the circuit and how
the parameters $\mathbf{\omega}_k$ parameterize the circuit will also
have an impact on the performance of the algorithm, and illustrate the
large flexibility available for designing QKSs tailored to particular
datasets and applications.

The choice of distributions and the parameterization of the circuit
together implicitly define a kernel which allows for QKSs to be analyzed
as a standard kernel machine, as we describe later. The computation of
the kernel is not necessary for the use of the QKS, and in fact may
require exponentially large resources, but it may be helpful in
designing the circuit Ans\"atze.

Once we have encoded the data into control parameters, we are ready to
preprocess the data. Since the input data is encoded in circuit
parameters, the choice of input state is somewhat arbitrary. For
simplicity and without loss of generality, we choose the all-zeros state
$\ket{\Psi_{\text{in}}}=\ket{00\ldots}$. Since any other input state would be
generated by another quantum circuit, the composition of this circuit
with the QKS encoding would correspond to a different circuit Ansatz.

In order to postprocess the QKS output, we must also extract classical data
from the state. This is done by simply measuring the state in the
computational basis---again, without loss of generality, since a
basis transformation would simply translate into changing the circuit
Ansatz. The output of the measurements gives us classical
bits. We have some design freedom in choosing how to (classically) 
process these output bits into features. Under the LB rule, care 
should be taken in this choice such that nonlinear postprocessing is 
avoided. For this work, we will simply ``stack" all of the bits into 
a $q$-dimensional feature vector.

Contrary to the RKS approach, this feature mapping
is stochastic. Our proposal does not preclude averaging over many
shots of the same circuit, but the numerical studies described here use only
individual shots of each circuit. 

Once we have constructed our feature vectors, they are fed into a
classical machine learning algorithm, which under the LB rule, we take
to be linear (as is also the case in RKS).

It is well-known in machine learning that transforming
data into a higher-dimensional feature space can be useful. 
There are two strategies to generate higher-dimensional features using QKS: entangling
more and more qubits, or generating more and more random circuits.
The first strategy leads straightforwardly to a quantum advantage
argument if the parameterized circuits used are hard to
simulate~\cite{AA2011,FH2016,BMS2016,BMS2017}. However, large, monolithic circuits
may also require very low error rates. The second strategy is more
readily scaled in NISQ devices, and it simply requires running $E$
fixed circuits, which we call {\em episodes}, to obtain a feature
vector that is $(E\times q)$-dimensional for $q$ qubits. We expect $D$
parameters in RKS should be roughly equivalent to $E\times q$ parameters in QKS,
but we do not have formal results that guarantee this correspondence.

\begin{figure}[t]
  \includegraphics[width=\columnwidth]{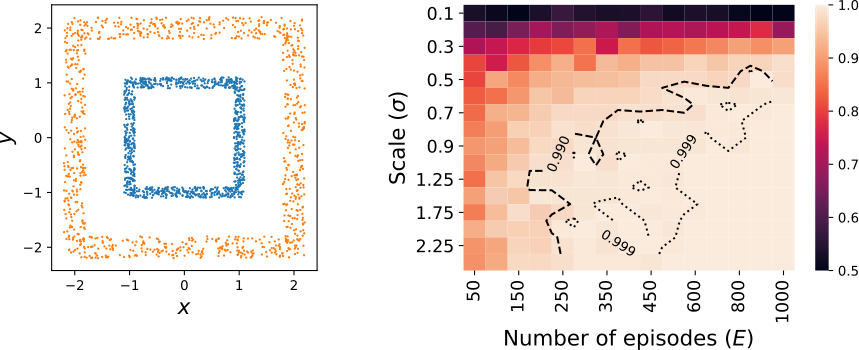}
  \caption{(a)~Synthetic ``picture frames'' dataset.
    (b)~Performance of the QKS classifier combined with logistic regression.
    We show the result of optimizing the performance as a function of the 
    hyperparameters $\sigma$ and $E$. The contours separate orders of 
    magnitudes in error rate. We see that optimal performance 
    (with a test accuracy of $>99.9\%$) is achieved with $\sigma \approx 1$.
    \label{fig:2d_frames}}
\end{figure}

\paragraph{A synthetic example---} As an example to demonstrate the effectiveness of
QKS, we incorporate it into a standard binary classification problem.
As our classical, linear baseline, we use the logistic regression (LR)
classifier provided by the \verb|scikit-learn| package.  As a first data set, we
choose the synthetic ``picture frames" dataset shown in
Fig.~\ref{fig:2d_frames}.  The dataset was chosen to have two classes
that are not separable by a linear boundary. The training set
contained $M=1600$ two-dimensional points, $800$ for each class.  The
classification accuracy was tested using a different set of $400$ points
arranged in a similar configuration.

We coded the algorithm using the
pyQuil\textsuperscript{\textregistered} Python
package~\cite{pyQuil,Smith2016} and executed it on the
Rigetti~QVM\texttrademark{}, available through the Forest
platform~\cite{Forest}. The QVM is a
high-performance quantum simulator written in ANSI Common
Lisp~\cite{ANSI:1996:ANS}. In order to run the numerical experiments
in conjunction with post-processing software in Python, the QVM was
extended with a new entry-point to allow high-speed execution of a
large number of episodes (on the order of $10^4$) for a given
circuit~Ansatz and input $\mathbf{u}$.  In particular, the QVM was extended so that a
template Quil~\cite{Smith2016} program defined with the
\verb|DEFCIRCUIT| facility could be supplied along with a collection
of \verb|DEFCIRCUIT| parameter tuples. The QVM reads these
parameter tuples, fills them into the supplied program in constant
time, and executes the resulting program, all while eliminating
unnecessary memory access and allocations. This modification to the
QVM was made possible using Quil's hybrid classical/quantum memory
model. See the Appendix (e.g., Fig.~\ref{fig:p16}) for examples of circuit
Ans\"atze written in Quil.

Applying the baseline LR algorithm to the picture frame dataset
yielded a classification accuracy of approximately $50\%$, meaning it
performs no better than randomly assigning classes to each point.
We then used the QKS construction, using the circuit shown in
Fig.~\ref{fig:2q_cnot}. For the data presented here, we used split
encoding (defined above) with $q=p=2$ and $r=1$, and optimized over
the number of episodes $E$ and the parameter $\sigma$ used in the random encoding.
The best classification accuracy achieved was $>99.9\%$, a remarkable
performance lift over the linear baseline, illustrating
the power of QKS (see Fig.~\ref{fig:2d_frames}).

\paragraph{A real-world example---} While this synthetic example
illustrates the computational power provided by the QKS, it is
interesting to consider a less structured classification problem
originating in the real world: discriminating hand-written digits from
the MNIST dataset~\cite{LBBH1998}. This
dataset is a well-known benchmark in machine learning. While
it is a multiclass problem, we choose to focus on classifying two
digits that are difficult to distinguishing using LR: ``3'' and
``5''. The classification accuracy we obtain with LR is $95.9\%$,
which will serve as our linear baseline.

\begin{figure}[t]
\includegraphics[width=0.55\columnwidth]{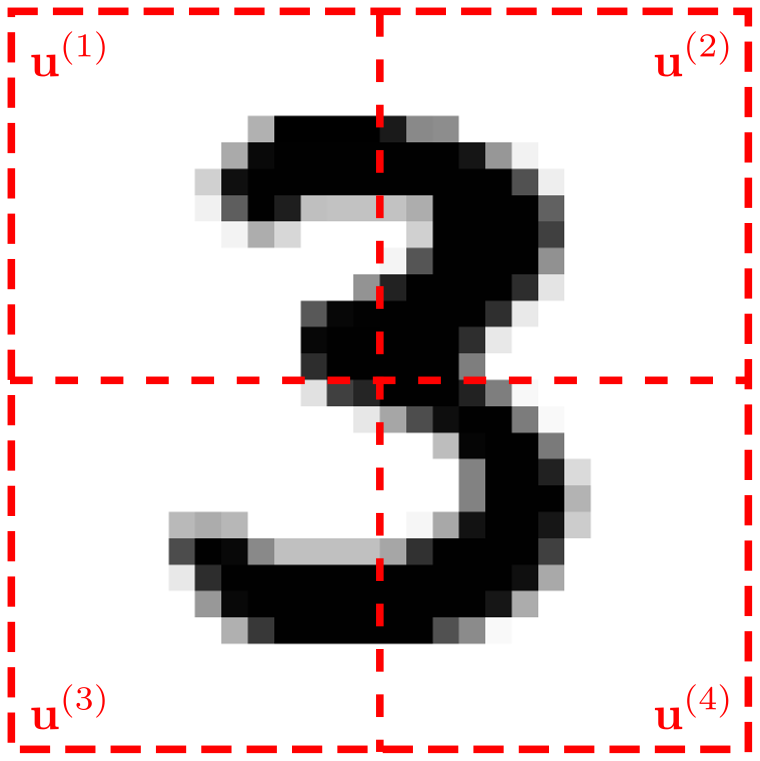}
\caption{
  A four-qubit partitioning of an example MNIST digit. Each partition, called a \emph{tile},
  corresponds to disjoint collection of components of the input vector $\mathbf{u}$,
  i.e., $\mathbf{u}$ is some permutation of the vector
  $(\textbf{u}^{(1)}, \textbf{u}^{(2)}, \textbf{u}^{(3)}, \textbf{u}^{(4)})^{\transpose}\in\mathbb{R}^{784}$.
  The exact permutation is encoded in the choice of nonvanishing values of the matrix $\mathbf{\Omega}$.
  \label{fig:partition}
}
\end{figure}

The MNIST dataset has a much higher dimensionality than the previous
example. Each digit is a $(28\times 28)$-pixel 8-bit grayscale image,
so care must be taken to encode the data into a small number of
qubits. A standard first step is to vectorize the image, by stacking
the columns of the image into a $p=784$ dimensional vector. We use a
slightly modified approach intended to preserve more of the spatial
structure of the image. After standardizing the
image~\footnote{\textit{Standardization} or \textit{$z$-normalization}
  of a set $X\subset\mathbb{R}$ is the pointwise map
  $x\mapsto (x-\mathrm{E}[X])/\operatorname{Var}(X)^2$.}, to run MNIST
on a $q$-qubit processor, we first split each image into $q$
rectangular tiles, and construct fixed-depth circuit Ans\"atze where
only single-qubit gates have parameterized rotations (see
Fig.~\ref{fig:partition}). The encoding vectors $\mathbf{\omega}_k$
are then chosen to have blocks of $r=p/q$ nonzero elements that select
out values of only one tile per gate parameter.

With this encoding, we have simulated the performance of QKS on the
$(3,5)$-MNIST dataset for different numbers of qubits. The best error rate is
$1.4\%$, which is a reduction of the error rate by more than a factor
of 2 compared to the linear baseline.

\begin{figure}[t]
    \centering
    \includegraphics[width=0.9 \columnwidth]{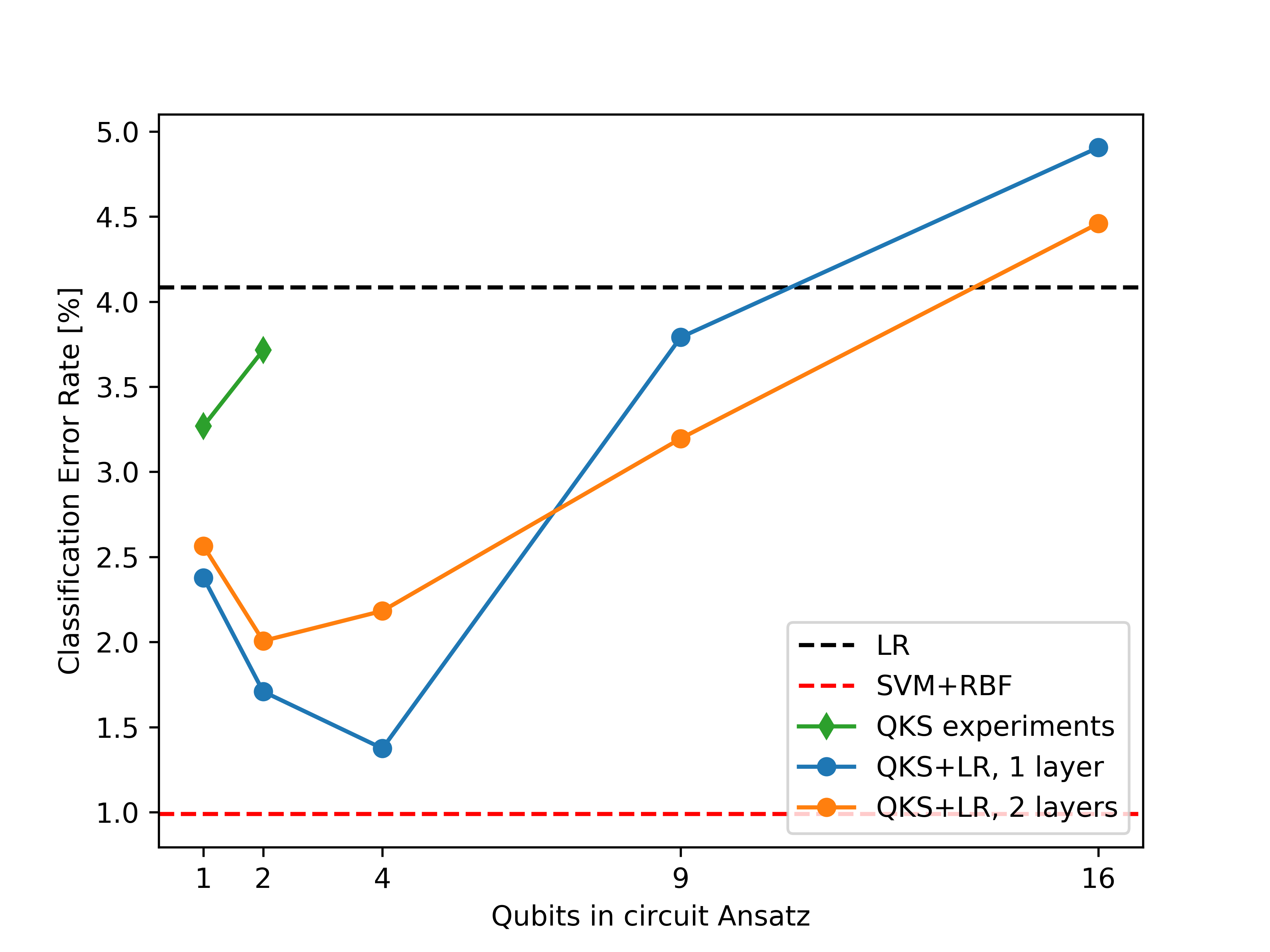}
    \caption{Scaling of the error rate
      classifying the $(3,5)$-MNIST dataset using QKS combined with
      logistic regression (LR), as a function of the
      number of qubits. We include both QVM and experimental QPU results. As a reference, we include the performance of
      LR on its own (our linear baseline) and the
      performance of a nonlinear classifier built out of a support
      vector machine (SVM) with a radial basis function (RBF) kernel. 
      Details of the circuit Ans\"atze can be found in
      Fig.~\ref{fig:2q_cnot} and the appendix. }
    \label{fig:mnist_scaling}
\end{figure}

In Fig.~\ref{fig:mnist_scaling} we plot the minimum error rate for
classifying the $(3,5)$-MNIST dataset using QKS for different numbers
of qubits. Each point corresponds to the minimum error observed after optimizing the hyperparameters $\sigma$ and $E$, much like what is shown in Fig.~\ref{fig:2d_frames}. The maximum number of episodes used was 20,000.  Comparing the results for different numbers of qubits, there is a clear minimum in the error rate in the range 2 to 4 qubits. For more qubits, the error rate increases again. There are a number of possible explanations for this behavior. One possibility is simply 
that the particular circuits chosen (including the \gatename{CNOT} networks) may be suboptimal for this task. Another possibility is that the MNIST data set 
is insufficiently large to properly train the larger number of parameters 
in the larger circuits, leading to overfitting. These and other hypotheses 
will be explored in future work.

It is also important to point out that quantum coherence does not play
a role in ``single layer'' circuits, as the same output mapping can be
implemented using purely classical stochastic processes (i.e., the
circuits are efficiently simulatable, in the weak sense~\cite{Nest}).
We are able to show, however, that by increasing the number of layers
of the circuit Ans\"atze (each layer using independently chosen linear
encodings) we are able to maintain similar performance (see
Fig.~\ref{fig:mnist_scaling}). One may also consider other Ans\"atze
based on circuits that are conjectured to be hard to
simulate~\cite{AA2011,FH2016,BMS2016,BMS2017}.

\paragraph{Experiments---}
A powerful feature of QKS is that specifying the form of $\mathbf{\Omega}$ allows a form of built-in 
compression, enabling us to classify full-sized MNIST images on small circuits. To further demonstrate this 
power and flexibility, we ran the classification problems we discussed here---picture frames and the (3,5) 
subset of MNIST---on one and two qubit circuits using a Rigetti QPU, \textit{i.e.}, on real quantum hardware. 
To our knowledge this is this is the first time image classification of MNIST images has been performed on a 
quantum computer.

These experiments were run on a device with 8 qubits arranged in a loop~\cite{CDRS+2018,ROTS+2018}. The 
\gatename{CNOT} 
gate was implemented as a \gatename{CZ} operation that native to our architecture~\cite{DSSR2018}, 
preceded and followed by Hadamard operations on one of the qubits~\cite{BBCD+95}. Average gate 
fidelities~\cite{Nielsen2002} for the CZ gate varied between 81\% and 91\% for the duration of these
experiments, which ranged from 1.5h for the picture frames, to 12h and 34h for the 1 and qubit 
ansatze applied to the (3,5)-MNIST dataset (gates were not recalibrated during the experiments, 
only between them). Different episodes were measured at a rate of roughly 400~Hz, which is largely a 
consequence of the choice to take a single of each episode for the feature generation (much higher 
rates can be obtained for multiple shots of a fixed circuit, but that feature is not exploited here).

Using the 2 qubit ansatz from Fig.~\ref{fig:2q_cnot}(a) on the picture frame data set, with 1000 episodes and
$\sigma=1.0$, we achieved 100\% accuracy on a test set of 400 examples and training set of 1600 examples, even 
in the presence of relatively noisy entangling gates.

We considered both the 1 qubit and the 2 qubit ansatze for the (3,5)-MNIST dataset (see the Appendix for 
details of the 1 qubit circuit). We did not use larger ansatze due to the connectivity limitations of the 
device used. We used 50\% of the MNIST dataset, and did not optimize the hyperparameters 
using the experimental setup, using instead simulations with the QVM to choose the hyperparameters. The 
motivation for these choices was to reduce the runtime
of the experiment. The experimental results are shown, along with the simulation results, in 
Fig.~\ref{fig:mnist_scaling}. For a one qubit circuit with $\sigma = 0.05$ and $E=10,000$, we find an error 
rate of $3.3\%$. Even for this simplest circuit, we find QKS provides a performance lift over a linear suport 
vector machine. Although, the error rate is somewhat higher than the best QVM result, it compares 
remarkably well to the one qubit QVM result with the same hyperparamters, which is an error rate of $3.2\%$. 
Implementing the two qubit \gatename{CNOT} circuit shown in Fig \ref{fig:2q_cnot} with $\sigma = 0.05$ and 
$E=8,900$, we find an error rate of $3.7\%$. Again, this shows a clear lift despite the use of real, noisy 
quantum hardware. 

\paragraph{The implied kernels---} The
random sampling of nonlinear feature maps across different episodes
can be connected to the use of an implicit kernel
function~\cite{RR2008,RR2008b,RR2009}. Formally, the kernel is the inner
product between input vectors after their nonlinear mapping by the
kitchen sinks. Informally, the kernel function $k(\mathbf{u},\mathbf{v})$ of two input
vectors $\mathbf{u}$ and $\mathbf{v}$ expresses the similarity between
these inputs. Even though the random and quantum kitchen sinks do not
explicitly use the kernel, it is instructive to calculate the implicit
kernel associated with our circuits, as it can point to better ways to
build circuit Ans{\"a}tze, and consider the effect of noise.

We compute the implicit kernel by evaluating the inner product of two
binary feature vectors sampled using a QKS circuit. Let
$\mathbf{b}_{e, \mathbf{u}}(\pmb{\theta}_e)$ and
$\mathbf{b}_{e, \mathbf{v}}(\pmb{\theta}_e)$ denote the vectorized
output of a single episode $e$ with the random parameters
$\pmb\theta_e$ on the inputs $\mathbf{u}$ and $\mathbf{v}$. The inner
product of the total feature vector can then be computed as
\begin{equation*}
    \tilde{k}(\mathbf{u},\mathbf{v}) =  \frac{1}{E} \sum_{e=1}^{E} \mathbf{b}_{e,\mathbf{u}}(\pmb{\theta}_e) \cdot \mathbf{b}_{e, \mathbf{v}}(\pmb{\theta}_e)
\end{equation*}
where we have added the normalization by $E$.

The quantities $\mathbf{b}_{e, \mathbf{v}}(\mathbf{\theta}_e)$ are
random variables with bit-string values $z \in \{0, 1\}^q$. The
probability of a given outcome $z$ is
$p_{e,\mathbf{u}}^{(z)}=\left|\bra{z}U(\mathbf{u},\mathbf{\theta}_e)\ket{\Psi_{\text{in}}}\right|^2$
where $U(\mathbf{u},\mathbf{\theta}_e)$ is the unitary transformation
realized by the QKS circuit.  We then find
\begin{equation*}
   \langle \mathbf{b}_{e,\mathbf{u}} \cdot \mathbf{b}_{e,\mathbf{v}} \rangle = \sum_{s=0}^q s P(\mathbf{b}_{e, \mathbf{u}} \cdot \mathbf{b}_{e,\mathbf{v}} = s) = \mathbf{p}^{\transpose}_{e,\mathbf{u}}\,\mathbf{S}\,\mathbf{p}_{e,\mathbf{v}}
\end{equation*}
where the matrix $\mathbf{S}$ contains the inner product of the bit
strings $z$ and $z'$, and the vector $\mathbf{p}_{e,\mathbf{u}}$ the outcome
probabilities, both indexed by $z$.

We now note that since the parameters $\pmb{\theta}_e$ are drawn from
a classical probability distribution $P(\pmb{\theta})$, we can view
the sum $\tilde{k}(u,v)$ as a Monte Carlo estimator.  In the limit of
an infinite number of episodes ($E\rightarrow\infty$), the kernel then
approaches the form
\begin{equation}\label{KernelInt}
    k(\mathbf{u},\mathbf{v}) =  \int\text{d}\pmb{\theta}~P(\pmb{\theta})~\mathbf{p}^{\transpose}_{\mathbf{u}}(\pmb\theta)\,\mathbf{S}\,\mathbf{p}_{\mathbf{v}}(\pmb\theta).
\end{equation}

Using this result, we can, for instance, calculate the implicit kernel
for the circuit in Fig.~\ref{fig:2q_cnot}(a). To do so, we specify
that the values of the matrix $\Omega$ are drawn from a normal
distribution $\mathcal{N}(0,\sigma^2)$ and that the elements of the
bias vector $\pmb\beta_e$ are drawn from a uniform distribution
$\mathcal{U}(0, 2\pi)$. Using \eqref{KernelInt}, we then find the
implicit kernel as
\begin{equation}\label{eq:2q_cnot_kernel}
    k(\mathbf{u},\mathbf{v}) = \frac{1}{2} + \frac{1}{8} \textrm{e}^{-\frac{1}{2}\sigma^2 \|{\bf u}^{(1)}-{\bf v}^{(1)}\|_2^2} + \frac{1}{16} \textrm{e}^{-\frac{1}{2}\sigma^2 \|\mathbf{u}-\mathbf{v}\|_2^2},
\end{equation}
where ${\bf u}^{(i)}$ (${\bf v}^{(i)}$) is the $i$\textsuperscript{th} tile
(out of 2) of the input data vector $\mathbf{u}$
($\mathbf{v}$)~\footnote{For the picture frame data set, this is just
  the $i$\textsuperscript{th} vector component.}.  We see that the last term here is a
radial basis function (RBF) kernel that is standard in machine
learning. There are additional components, including a constant
term. The second term depends only on part of the data. Similar
calculations can be performed for the other circuit Ans\"atze, and
again we find multiple terms that depend on different subsets of the
data. One can imagine optimizing the \gatename{CNOT} network to
maximize sensitivity to the most relevant subsets of the data, but we
do not explore the possibility here.

Interestingly enough, not all circuit Ans\"atze lead to a useful
kernel. For instance, circuit Fig.~\ref{fig:2q_cnot}(b) seems similar
to the just-analyzed circuit. However, if we calculate the implied
kernel of this circuit, we find the constant function
$k(\mathbf{u}, \mathbf{v}) = 1/2$, independent of the input vectors
$\mathbf{u}$ and $\mathbf{v}$. This suggests that this circuit should
have no discrimination power and, in fact, our numerical results
confirm this.

We see that QKS provides a rich structure to construct implicit
kernels, with not only the choice of circuit, but the choice of
encoding, choice of decoding, and choice of probability distributions
shaping the kernel in understandable ways.

\paragraph{Discussion---} 
For context, we can compare our experimental MNIST results to simulation results based on other algorithms.  For instance, ref.~\cite{Huggins2018} simulates the same $(3,5)$-MNIST classification problem, using images downsampled to $8\times8$ pixels on much wider and deeper networks.  Even with noiseless circuits, they achieve an error rate of only $12.4\%$, much higher than our \textit{experimental} error rates.

While ref.~\cite{Havlicek2018} focuses on a variational algorithm, it also studies a second, open-loop algorithm.  This hybrid algorithm uses the QPU to directly estimate a kernel matrix, which can then be used in standard, classical kernel algorithms.  We can compare the quantum resources required for the training phase in this approach to QKS, which uses an explicit transformation instead of a kernel function or matrix. Ref.~\cite{Havlicek2018} finds that, in order to estimate the kernel matrix with an operator error of $\epsilon$, the number of calls to the QPU required is $O(\epsilon^{-2} M^4)$, where we recall that $M$ is the number of training examples. By comparison, QKS requires $R = E M$ calls to the QPU.  While we have not derived complexity bounds for QKS, we find numerically that the number of episodes, $E$, required is the same order as for RKS. For RKS, to estimate our classification function $g$ with error $\epsilon$ requires $E \sim O(\epsilon^{-2})$ episodes. Using this bound, we then find the number of QPU calls required for QKS to be $R \sim O(\epsilon^{-2} M)$, which is a substantial improvement over the result of ref.~\cite{Havlicek2018}.  We recall that RKS was developed to improve on the complexity of classical kernel algorithms, and it seems that QKS inherits that improvement.

\paragraph{Conclusions---} 
We have described how random quantum circuits can be used to transform classical data in a highly 
nonlinear yet flexible manner, similar to the random kitchen sinks technique from
classical machine learning. These transformations, which we dub {\em quantum
  kitchen sinks}, can be used to enhance classical machine learning
algorithms. We illustrated this enhancement by showing that the
accuracy of a logistic regression classifier can be boosted from $50\%$
to $>99.9\%$ in low-dimensional synthetic datasets, and from $95.9\%$ to
$98.6\%$ in a high-dimensional dataset consisting of the
hand-written ``3'' and ``5'' digits of the MNIST database. In all these
examples, this can be achieved with as few as four qubits. We also presented experimental
results using 1 and 2 qubits that showed similar performance, with accuracies
of $100\%$ for the picture frames, and $>96\%$ for the subset of the MNIST database discussed 
here. This outperforms near term proposals using tensor networks~\cite{Huggins2018}, and 
has similar performance to other algorithmic proposals~\cite{KL18}, with the advantage that QKS
uses much lower depth circuits and therefore can tolerate much higher error rates in the experiments.
Future work will focus on exploring different circuit Ans\"atze, and developing a
better understanding of the performance of this technique.

\paragraph{Contributions---} CMW proposed the original concept of extending random 
kitchen sinks to quantum kitchen sinks, CMW and JO developed the
theory and prototyped the numerical analysis. JO, NT, and RSS
developed the scalable analysis for larger datasets. MSA contributed
to the analysis of the classification error rates.  GEC proposed the
LB rule. AMP collected and analyzed experimental data, collaborating
with PJK and SH to build and optimize the QPU data collection
framework.  MPS supervised and coordinated the effort. CMW, JO, NT,
RSS, and MPS wrote the manuscript.

\paragraph{Acknowledgements---} We acknowledge helpful discussions
with Matthew Harrigan.

\bibliography{qks}

\pagebreak
\appendix

\section{Picture Frames Dataset}

The picture frames dataset (Fig.~\ref{fig:2d_frames}) was chosen to
have a nontrivial shape and such that the two classes were not
linearly separable.  The smaller (red) square has a side length of 2
with points uniformly distributed in a region 0.1 around the
average. The larger (blue) square has a side length of 4 with points
uniformly distributed in a region 0.2 around the average.

\section{Parameterized programs for circuit Ans\"atze}\label{sec:defcircs}
Figs.~\ref{fig:p4}, \ref{fig:p8}, and \ref{fig:p16} define circuit Ans\"atze for 4, 8, and 16 qubits respectively using the \verb|DEFCIRCUIT| facility in Quil~\cite{Smith2016}. \verb|DEFCIRCUIT| defines a template which can be filled in via the \verb|%|-prefixed parameters.

\begin{figure}[h]
\begin{verbatim}
DEFCIRCUIT P4(%x0,%x1,%x2,%x3):
    RX(%x0) 0
    RX(%x1) 1
    RX(%x2) 2
    RX(%x3) 3
    CNOT 0 2
    CNOT 1 3
    CNOT 0 1
    CNOT 2 3
\end{verbatim}
\caption{A four-qubit QKS Ansatz written using a Quil \texttt{DEFCIRCUIT}.\label{fig:p4}}
\end{figure}

\begin{figure}[h]
\begin{verbatim}
DEFCIRCUIT P9(%x0,%x1,%x2,%x3,%x4,%x5,%x6,%x7,%x8):
    RX(%x0) 0
    RX(%x1) 1
    RX(%x2) 2
    RX(%x3) 3
    RX(%x4) 4
    RX(%x5) 5
    RX(%x6) 6
    RX(%x7) 7
    RX(%x8) 8
    CNOT 0 3
    CNOT 1 4
    CNOT 2 5
    CNOT 3 6
    CNOT 0 1
    CNOT 3 4
    CNOT 5 8
    CNOT 6 7
    CNOT 1 2
    CNOT 4 7
    CNOT 4 5
    CNOT 7 8
\end{verbatim}
\caption{A nine-qubit QKS Ansatz written in Quil.\label{fig:p8}}
\end{figure}
\newpage
\begin{figure}[h]
\begin{verbatim}
DEFCIRCUIT P16(%x0,%x1, ..., %x14,%x15):  # params elided
    RX(%x0) 0
    RX(%x1) 1
    RX(%x2) 2
    RX(%x3) 3
    RX(%x4) 4
    RX(%x5) 5
    RX(%x6) 6
    RX(%x7) 7
    RX(%x8) 8
    RX(%x9) 9
    RX(%x10) 10
    RX(%x11) 11
    RX(%x12) 12
    RX(%x13) 13
    RX(%x14) 14
    RX(%x15) 15
    CNOT 0 4
    CNOT 1 5
    CNOT 2 6
    CNOT 3 7
    CNOT 8 12
    CNOT 9 13
    CNOT 10 14
    CNOT 11 15
    CNOT 0 1
    CNOT 2 3
    CNOT 4 5
    CNOT 6 7
    CNOT 8 9
    CNOT 10 11
    CNOT 12 13
    CNOT 14 15
    CNOT 1 2
    CNOT 4 8
    CNOT 5 9
    CNOT 6 10
    CNOT 5 6
    CNOT 7 11
    CNOT 9 10
    CNOT 13 14
\end{verbatim}
\caption{A 16-qubit QKS Ansatz written in Quil.\label{fig:p16}}
\end{figure}


\end{document}